# Synthesis and electrical properties of fullerene-based molecular junctions on silicon substrate

D. Guérin, S. Lenfant, S. Godey, D. Vuillaume

*Molecular Nanostuctures and Devices group, Institut d'Electronique, Micro-électronique et Nanotechnologie , CNRS, University of Lille, BP60069, avenue Poincaré, F-59652 cedex, Villeneuve d'Ascq (France)*

We report the synthesis and the electrical properties of fullerene-based molecular junctions on silicon substrate in which the highly π-conjugated molecule $C_{60}$ (π quantum well) is isolated from the electrodes by alkyl chains (σ tunnel barriers). Initially, the $Si/SiO_2/\sigma C_{60}$ architecture was prepared either by sequential synthesis (3 different routes) or by direct grafting of the presynthesized $C_{60}$-σ-$Si(OEt)_3$ molecule. We described the chemical synthesis of these routes and the physico-chemical properties of the molecular monolayers. Then, the second σ tunnel barrier was added on the $Si/SiO_2/\sigma C_{60}$ junction by applying a hanging mercury drop electrode thiolated with an alkanethiol monolayer. We compared the electronic transport properties of the $Si/SiO_2/\sigma C_{60}//Hg$ and $Si/SiO_2/\sigma C_{60}//\sigma Hg$ molecular junctions, and we demonstrated by transition voltage spectroscopy that the fullerene LUMO - metal Fermi energy offset can be tailored from ~ 0.2 eV to ~ 1 eV by changing the length of the alkyl chain between the $C_{60}$ core and the Hg metal electrode (i. e. from direct $C_{60}//Hg$ contact to 14 carbon atoms tunnel barrier).

Surface functionalization utilizing buckminsterfullerene $C_{60}$ is of great interest, owing to the possibility of transferring the exceptional electronic (superconductivity, switching, memory), optoelectronic and magnetic properties[1] of this molecule on the surface of bulk materials. The electronic structure of this highly π-conjugated molecule with relatively high electron affinity (2.65 eV)[2] and low band gap (1.6 eV)[3] makes it a strong electron-accepting group coveted in numerous applications, particularly in nanoelectronics. For instance, high electron field effect mobility (0.2 $cm^2$/Vs) in n-channel Organic Thin Film Transistors (OTFT) has been reported for $C_{60}$ film on HMDS modified $SiO_2$ as gate dielectic.[4] Resonant tunneling effect has been observed by Grüter et al.[5] in a thiolated $C_{60}$ molecular junction prepared by break junction method between gold electrodes. Bistable and negative differential resistance (NDR) effects have been observed in single layer devices made from fullerenes mixed with polystyrene and sandwiched between two Al electrodes.[6] It has been recently reported several $C_{60}$-based unimolecular rectifiers with high rectification ratios up to 158 at 3 V.[7]

To our knowledge, all the previously reported studies in the field of nanoelectronics on $C_{60}$-based molecular junctions concerned $C_{60}$ deposited by vacuum evaporation[8], by the Langmuir-Blodgett technique[9], $C_{60}$ inserted in mechanically break junction[5], e-beam lithographed nano-gap[10], or scanning tunneling microscopy (STM)[11] test-beds. Here we report on molecular junctions on silicon based on self-assembled monolayers (SAMs) including $C_{60}$ molecules.

Electronic coupling between molecules and electrodes is a key point impacting the electron transport in molecular junctions. For instance, many groups have reported on the influence of the chemical nature of the chemical linker (e.g. thiol, selenol, amine, phosphine…) between the molecule and a metal electrode.[12] In SAM rectifying diodes based on the silicon/σ-π/metal architecture, Fermi-level pinning at the π group/metal interface has been identified as the main responsible for the absence of dependence of the rectification effect as a function of the chemical nature of the π groups, although these groups have significant differences in their electronic structure (i.e. different HOMO and LUMO energy levels).[13] This effect has been interpreted as resulting from orbital overlaps between metal and π groups which create new chemically-induced gap states at the metal/organic interface.

Here, we report results on the dependence of this orbital coupling effect as function of the length of the saturated spacer (σ chain) between the $C_{60}$ core and the metal electrode. To this end, we compared the electrical properties of two molecular architectures: $Si/SiO_2/\sigma C_{60}//Hg$ and $Si/SiO_2/\sigma C_{60}//\sigma Hg$ junctions. In the first architecture, π-conjugated group was directly in contact with the metal whereas in the second one, $C_{60}$ core was isolated between two alkyl chains (called σ tunnel barriers). Natively oxidized silicon (~10 Å) was used. Metal electrode was deposited on the $Si/SiO_2/\sigma C_{60}$ molecular junctions by applying hanging mercury drop on the monolayer. Moreover, by treating the Hg drop electrode with alkylthiols SAMs of different lengths, the π goup was isolated from the metal by a second σ tunnel barrier. Electrical characterizations of the resulting $Si/SiO_2/\sigma C_{60}//Hg$ and $Si/SiO_2/\sigma C_{60}//\sigma Hg$ molecular devices were compared.

Many strategies have been developed in the literature to realize the self-assembly of fullerene and its derivatives on various surfaces.[14] Among them are two commonly used methods : *(a)* the sequential modification of the surface by adsorption of an organic monolayer bearing a terminal reactive functionality such as amine[15], azide[16], or azomethine ylide[17] group that forms coupling reaction with $C_{60}$, or *(b)* the direct assembly of prefunctionalized $C_{60}$ derivatives bearing suitable group that binds the surface. Trialkoxysilane groups $-Si(OR)_3$ are appropriate functions in the case of oxidized silicon surface since they strongly chemisorb forming dense and well-organized self assembled monolayers (SAMs).[18]

In the present work, we compared these two strategies (sequential and direct grafting) in order to prepare good quality $C_{60}$ monolayers on oxidized silicon. Physicochemical characterizations of SAMs by several analytical tools (contact angle measurement, ellipsometry, FTIR spectroscopy, X photoelectron spectroscopy, cyclic voltammetry) enabled to select the best molecular junctions for their electrical characterization.

**1- Synthesis of $Si/SiO_2/\sigma C_{60}$ molecular junction by sequential routes.**

We chose 3 sequential methods (schema 1) involving reactions of buckyball with SAMs bearing 3 different reactive end-groups. The first two selected sequential routes were based on the well-known 1,3-dipolar cycloaddition reaction of azomethine ylides to fullerene (also called Prato reaction).[19] In the presence of $C_{60}$ the ylide adds to buckyball yielding a monolayer in which a pyrrolidine ring is fused to $C_{60}$ (i.e. fulleropyrrolidine). Azomethine ylides are generated *in situ* either (sequential route 1) by a decarboxylation route[17a] from aldehyde terminated SAM and N-methylglycine at 180°C or (sequential route 2) by thermal ring-opening of carbomethoxyaziridine-terminated SAM.[20] The last sequential route brought into play azide function as a reactive group in the presence of $C_{60}$ (sequential route 3) giving rise to fulleroaziridine adduct by thermolysis (and probably also azafulleroid and polyazafulleroid derivatives not represented on schema 1).[21]

Between $C_{60}$ core and silicon substrate, we voluntarily chose small alkyl spacers of only 3 carbon atoms playing the role of a small tunnel barrier (noted σ on schema 1) in order to not restrict current density through the molecular junction. Precursor SAMs bearing aldehyde, carbomethoxyaziridine and azide functions were prepared by classical silanization reaction. The details of these syntheses and their physicochemical characterizations are given in the supplementary information section.

**$C_{60}$ grafting by sequential route 1.**

Among the several successful $C_{60}$ functionalization methodologies reported in the past years, the Prato reaction is certainly the most widely used. Therefore, it seemed of interest to undertake grafting of $C_{60}$ on the CHO-terminated SAM by the Prato reaction. Unlike results obtained by Cattaruzza et al.[17] from CHO-terminated long aliphatic chains on Si-H surface, Prato reaction was unsatisfactory with our CHO-terminated SAM in the presence of $C_{60}$ and N-methylglycine. Indeed, solution turned dark red during the grafting reaction and by products were observed in the medium by thin layer chromatography. We did not try to isolate these by-products that probably resulted from side reactions of N-methylglycine with fullerene. Literature results show that yields of Prato reactions are generally low in homogeneous phase (~20-30%). Even if the film thickness seemed correct (16 Å), the static water contact angle (44°) was too low compared to expected value[15b,17,22] ranging from 65° to 76° for common $C_{60}$ monolayers (table 1).

**$C_{60}$ grafting by sequential route 2.**

Another option to generate azomethine ylide is the thermal ring-opening of carbomethoxyaziridine. Carbomethoxyaziridine functionalized with trialkoxysilane group had been already employed by Bianco et al. to synthesize a fullerene derivative bearing -$Si(OEt)_3$ group which was then used to functionalize HPLC silica gel.[14] To our knowledge, this reaction had never been realized by sequential surface chemistry on $SiO_2$ surface. We thus realized thermolysis of carbomethoxyaziridine-terminated SAM in the presence of $C_{60}$. Let us precise that such a reaction required heating the medium at least at 80°C for several hours and under these conditions the least oxidant or water traces damaged the SAM. We observed that presence of a water trap (freshly activated molecular sieves) in the medium dramatically reduced this problem but the reproducibility of the reaction remained low.

Static water contact angle (71°) was in agreement with literature values[15b,17,22] but thickness measurements (ranging from 15.4 Å to 44 Å) varied from sample to sample (table1). For most samples, $C_{60}$ multilayers were obtained suggesting strong physisorption of fullerene aggregates on the substrate that intensive cleaning by sonication in different solvents could not remove. Fullerene aggregation is an often observed phenomenon in surface functionalization by $C_{60}$.[23]

We wanted to confirm the presence of the electroactive specie $C_{60}$ by cyclic voltammetry. Figure 1 shows typical voltammograms obtained on $C_{60}$ monolayers and multilayers used as working electrodes. Depending on the number of fullerene layers, a more or less well-defined irreversible wave was observed in the cathodic region. Even if peak intensities were quite low due to the presence of resistive $SiO_2$ layer (9.7 Å), the reduction potential at approximately -1 V/SCE was close to that observed for other SAMs containing $C_{60}$.[17a,15b] As a reference, a saturated solution of $C_{60}$ in toluene/$CH_2Cl_2$ presented 3 reduction peaks at -1.01, -1.31 and -1.74 V/SCE, respectively (Cf. sequential route 3, fig. 5). These results tend to confirm the presence of fullerene on the oxidized silicon surface but it should be admitted that the $C_{60}$ grafting reaction is quite difficult to control and to reproduce by this method.

## $C_{60}$ grafting by sequential route 3.

The last sequential method brought into play thermolysis of $N_3$-terminated SAM in the presence of $C_{60}$. After this surface treatment, static water contact angle (table 1) slightly varied from 68° to 70° (typical literature value for $C_{60}$ film), while SAM thickness increased from 8 Å to 14 Å in agreement with theoretical MOPAC[24] value (14.6 Å). Infrared spectroscopy in ATR mode (Attenuated Total Reflexion), also confirmed disappearance of the characteristic absorption peak associated to -$N_3$ group at 2101 cm$^{-1}$ due to decomposition by thermolysis. Antisymmetric and symmetric -$CH_2$- stretches were observed at 2933 and 2869 cm$^{-1}$, respectively for $N_3$-terminated SAM, while they were observed at 2927 and 2860 cm$^{-1}$, respectively for $C_{60}$-terminated SAM (figure 2). In crystalline aliphatic chains, antisymmetric and symmetric -$CH_2$- vibrations occur at 2918 and 2851 cm$^{-1}$, respectively. It has been established that absorptions due to -$CH_2$- stretching vibrations shift to higher frequency when disorder increases (and thus the chain length decreases) in monolayers.[25] Here values observed for -$CH_2$- stretches are usual for small alkyl chains.

Vibration peaks related to $C_{60}$ (i.e. C=C double bonds) could not be assigned due to the complexity of spectra in the 700-1600 cm$^{-1}$ region (which is also the vibration region of Si-O bonds of the native silicon oxide). In a typical spectrum from literature[26], $C_{60}$ moiety vibration bands are observed at 526 (the most intense band), 630, 1180 and 1430 cm$^{-1}$.

XPS measurements were also achieved to follow sequential route 3. Figures 3(a) and 3(b) show N1s core level peak on $N_3$- and $C_{60}$-terminated SAMs respectively. As expected, the signature of azide group was evidenced on $N_3$- terminated SAM by 2 components at 401.2 eV and 404.9 eV with a characteristic area ratio of 2:1 respectively. The additional shoulder peak at lower binding energy (400.5 eV) was assigned to amine function resulting from azide decomposition under X ray radiation.[27] After $C_{60}$ grafting, as shown on figure 3(b), the two high binding energy peaks characteristic of azide completely disappeared and a new N1s component was found in the classical energy range of tertiary aliphatic amines at 399.9 eV in accordance with the formation of fulleroaziridine derivatives on the surface. The noticeable asymmetric shape of this N1s core level peak was due to the presence of another subjacent component at about 401.9 eV attributed to protonated amine[27] or to polynitrogen adducts such as triazolines[28] since multiple addition and rearrangement reactions between $C_{60}$ and azides may occur.[21] $C_{60}$ grafting was evidenced by the clear increase of C/N atomic ratio by a factor of about 5 between the $N_3$- and $C_{60}$- terminated SAMs (see figure 4). Nevertheless, the expected theoretical increase of C/N is 18, as C/N values should be 1 for the $N_3$- terminated SAM, and 18 for the $C_{60}$- terminated one. This last assumption relies on the steric hindrance difference between $C_{60}$ (~82 Å$^2$/molecule) and alkyl chain (~21 Å$^2$/molecule), which leads to a maximum of 1/4 chains linked to fullerene in a close-packed arrangement. Furthermore, here we assumed complete decomposition of unreacted azide groups into amine, imine or azo groups[29] by the loss of $N_2$ after the thermolysis reaction (i.e. by considering one nitrogen atom per alkyl chain). Discrepancies between expected and experimental C/N atomic ratios values arise mainly from secondary reactions as mentioned previously, but also from uncertainties for photoelectron attenuation lengths in SAM, from a possible additional carbon contamination and from high sensitiveness of azide group to X-ray.

Observation of fullerene redox system on $C_{60}$-terminated SAM was also realized by cyclic voltammetry. Reaction sequence was realized on n$^+$ doped oxidized silicon and also on ITO (less resistive substrate) in order to amplify the electrochemical signal. Reactivity of organosilanes was considered as similar on silicon oxide and indium tin oxide surfaces since both surfaces are made of -OH anchoring groups. We first determined (figure 5) the reduction potentials of a saturated solution of $C_{60}$ in toluene-$CH_2Cl_2$ (v/v 1:1) on the precursor $N_3$-terminated SAMs used as working electrode. Reduction peaks were observed at -1.01, -1.31 and -1.74 V/SCE. On the voltammogram of $N_3$-terminated SAM taken as a reference on figure 6, no electroactive specie was detected from 0.5 V/SCE to -1.75 V/SCE. After $C_{60}$ grafting on $N_3$-terminated SAMs, voltammograms (figure 6) clearly confirmed the presence of fullerene on $Si/SiO_2$ and ITO substrate. Indeed, a reduction wave ranging from -0.79 to -0.90 V/SCE (depending on the sample run) was measured on oxidized silicon. In the same way on ITO substrate, a first reduction peak was observed at -1.01 V/SCE while a second one was detected at ~-1.80V/SCE.

Surface topology of $N_3$- and $C_{60}$-terminated SAMs was compared by TM-AFM. As shown on figure 7, $N_3$-terminated SAM was very smooth with low rms roughness of 0.20 nm over 5×5 µm area close to supporting silicon wafer roughness. On the contrary, rms roughness of $C_{60}$-terminated SAM (1.07 nm) was more important with homogeneous repartition of $C_{60}$ aggregates (10-200 nm diameter, 1-30 nm thick) as previously observed in literature on various types of surfaces.[16,30] It was not possible to eliminate these $C_{60}$ clusters by sonication in different solvents indicating that aggregates were strongly chemisorbed on the monolayer.

Actually high temperature grafting for several hours appeared to be inappropriate for these low thickness monolayers contrary to certain successful examples from literature.[16-17] High temperature necessary for $C_{60}$ grafting damaged monolayers and most of theses reactions were difficult to reproduce.

**2- $Si/SiO_2/\sigma C_{60}$ molecular junction by direct grafting**

Given difficulties of reproducibility observed with previous sequential methods, we undertook direct grafting of preformed σ-$C_{60}$ molecule bearing trialkoxysilyl anchoring group on natively oxidized silicon. We carried out the synthesis of the fullerene derivative represented on schema 2 following the previously described Prato method[31] by reaction of N-methylglycine and triethoxysilylbutyraldehyde with $C_{60}$. In this fullerene derivative, -$Si(OEt)_3$ group is attached at 2-position of the pyrrolidine ring through a short propyl spacer. It was obtained in low yield (22 %) after purification by flash chromatography and was successfully characterized by $^1H$ and $^{13}C$ NMR (see the supplementary information section).

Contrary to preceding sequential methods (realized above 80°C and damaging the SAM), self assembly of the $C_{60}$ derivative on natively oxidized silicon substrate required 4 days in soft conditions (room temperature). Lower temperature should favor molecule organization in the monolayer. Obviously, to compare with preceding sequential routes, one cannot expect to get the same σ alkyl chains compactness at the bottom of the monolayer because of $C_{60}$ hindrance. Bulky molecular structures have an important impact on the surface coverage : generally, the bulkier the terminus, the lower the surface packing. Owing to that, in plane -Si-O-Si- polysiloxane reticulation should be also less favored. As indicated in table 1, SAM thickness measured by ellipsometry (14.4 Å) was in conformity with theoretical value (14.7 Å). Static water contact angle of $C_{60}$ monolayer was 70°, in agreement with literature.[17]

Chemical composition of $C_{60}$-functionalized SAM obtained by direct grafting was also evaluated by XPS analysis. The measured C/N atomic ratio was about 44, i.e. 3.5 times higher than in the case of sequential route 3, but the surface presented less carbon as can be evidenced for example by comparing relative intensities of C1s and Si2p on fig. 4(b) and fig.

8. This is consistent with the fact that the σ moiety is denser in the case of SAM obtained by sequential route 3, since it remains unreacted alkyl chains not linked to $C_{60}$. In the case of direct grafting, the high resolution spectra of N1s (see inset of fig. 8) and C1s (see fig. 9 solid line) presented less chemically distinct environments compared to sequential route 3 (fig. 3(b) and fig. 9 dashed line). The N1s core level, whose intensity was very low compared to sequential route 3, presented only one component. As for the N1s core level peak, C1s showed less asymmetry at higher binding energies (fig. 9) which means there were less carbon atoms involved in bonds with oxygen atoms. The oxygen concentration in the case of direct grafting mainly resulted from the silicon oxide of the substrate.

Figure 10 shows the voltammogram of $C_{60}$-terminated SAM prepared in the same conditions on two substrates: Si $n^+/SiO_2$ and ITO. As previously observed for sequential methods, a reduction peak was observed at -1.11 V/SCE confirming surface functionalization by $C_{60}$ electroactive species. This peak was better resolved on less resistive ITO substrate.

## 3- Electrical measurements of Si/SiO$_2$/σC$_{60}$//σHg molecular junction by direct grafting

We studied the electronic transport properties of the molecular junctions obtained by direct grafting since we have shown above that the SAMs formed by this route were more homogeneous and reproducible. To electrically characterize the molecular junction, we used an electric contact by mercury drop on top of the SAM.[32] This soft method avoids formation of short circuit through the monolayer. By comparison with direct evaporation of metal, the weak interactions between organic and metal avoid metal penetration in the organic material.[33] Another advantage of this method is the easy possibility to form a second σ tunnel barrier by grafting an alkylthiol monolayer on the mercury drop. Here, Hg drop was thiolated with two different alkyl chain lengths (C6 and C14 for short) by $CH_3(CH_2)_nSH$ alkylthiols (with n=5 and n=13 respectively). Then Hg drop (with or without the alkylthiol SAM) was gently contacted on the sample to define the two molecular junctions: Si/SiO$_2$/σC$_{60}$//Hg and Si/SiO$_2$/σC$_{60}$//σHg, respectively (figure 11).

Current-voltage characteristics were measured on these junctions (figure 12) for various lengths of the second tunnel barrier. We modified the length of the second barrier by changing the SAM grafted on the mercury drop. In figure 12, three junctions were compared i) without second tunnel barrier, i.e. the mercury drop directly in contact with $C_{60}$, ii) with a thin alkylthiol SAM (C6) grafted on mercury drop, and iii) with a thicker alkylthiol SAM (C14) grafted on mercury drop. As expected, the current decreased with the increase of the second tunnel barrier thickness. For example, at 1 V, the currents measured without second barrier, with a C6 SAM, and with a C14 SAM are $8.6 \times 10^{-8}$ A, $1.3 \times 10^{-10}$ A and $3.9 \times 10^{-12}$ A, respectively. As expected, in the framework of a non-resonant tunneling transport through the molecular junction, the current is exponentially dependent on the SAM thickness, and we deduce a tunnel attenuation factor of $0.6 \pm 0.2$ Å$^{-1}$. This value is as expected for a π conjugated molecule,[34] and similar to values reported for quite similar monolayer molecular junctions made of a short oligomer (oligophenylene) in-between two alkyl chains.[35]

We analyzed the I-V traces by transition voltage spectroscopy (TVS) method.[36] Recent works have shown that the TVS method gives a good determination of the position of the molecular orbitals with respect to the Fermi energy of the electrodes.[37,38] Figure 13 shows the plots of $\ln(I/V^2)$ vs. $1/V$ for the same I-V data as in figure 12, for the negative and positive voltages. The bias ($V_T$) at the minimum of $\ln(I/V^2)$ directly determines the energy offset $\Phi = e |V_T|$ between the Fermi energy in the electrodes and the frontier orbitals of the molecules.

The curves presented in the figure 13, show a clear difference for the $V_T$ as function of the distance between the $C_{60}$ and the Hg drop, i.e. as function of the thickness of the SAM grafted on Hg drop. Without the second tunnel barrier the $V_T$ values for the negative and positive part are respectively -0.22V and 0.24V. With the presence of the second tunnel barrier, the values for $V_T$ increase to -0.40 and 0.40 with C6 SAM, -0.66 and 0.59 with a C14 SAM. This typical example shows a clear tendency: the energy offset between the metal Fermi energy and the molecular orbital (whether it is the LUMO or HOMO will be discussed below) decreases when decreasing the coupling distance between the $C_{60}$ and the Hg electrode.

This tendency was confirmed by recording between 5 and 25 I-V traces for each junction. Figure 14 shows the dispersion of the measured current at 1 V probably due to inhomogeneities in the molecular organization of the SAM. Two peaks of $V_T$ are observed for each junction, corresponding to the negative and positive voltage. Without the second tunnel barrier the mean $V_T$ value is equal to -0.24 ± 0.09 (this latter value is the standard deviation of normal distribution shown in fig. 14) and 0.17 ± 0.10 for negative and positive bias, respectively. These $V_T$ values increase with the thickness of the second tunnel barrier to -0.35 ± 0.04 and 0.33 ± 0.03 for the C6 SAM, and to -0.97 ± 0.20 and 0.83 ± 0.18 for the C14 SAM. The slight difference of the values measured from the negative and positive biases can be due to i) the slight difference in the work function of the Si and Hg electrodes, ii) the slight geometrical asymmetry in the position of the $C_{60}$ core in between the two electrodes due to the different lengths of the σ spacers (note that the lowest asymmetry, -0.35 V and 0.33 V, is obtained when these spacers are not too different, that is with C3 and C6 chains, respectively).

From these energy levels determined by the TVS method, we deduced the energy diagram for each molecular junction (figure 15). We assumed that the $V_T$ value determined previously corresponds to the energy position of the LUMO with respect to the Fermi energy of the electrodes. This statement relies on two features: i) $C_{60}$ is an acceptor molecule with an electron affinity of about 2.7 eV and it is likely that charge transfer occurs via the LUMO, ii) scanning tunneling experiments (STS) on single $C_{60}$ deposited on H-passivated Si and slightly oxidized Si have clearly detected such a resonant transport through the LUMO and not the HOMO.[39,40] The mean value of 0.83-0.97 eV measured here when the $C_{60}$ is electronically decoupled from both the Si and the Hg electrodes with alkyl chains is consistent with the about 1 eV value reported with the STS experiments (in which the $C_{60}$ is also decoupled from the Si electrode by a thin $SiO_2$ layer and by the vacuum gap between the tip and the molecule)[3] and also in agreement with ab-initio calculations.[41] Our results clearly illustrate that increasing the electronic coupling between the molecule and the electrode, by decreasing the chain length of the alkanethiol molecule, results, as expected, in a lowering of the LUMO with respect to the Fermi energy of the electrode.[42]

4- Conclusion

The direct grafting of a presynthesized fullerosilane appeared to be the more efficient route to prepare $C_{60}$-based self assembled monolayer on natively oxidized silicon. Analysis of the I(V) characteristics of $Si/SiO_2/\sigma C_{60}//Hg$ and $Si/SiO_2/\sigma C_{60}//\sigma Hg$ devices showed electron transport through the LUMO of $C_{60}$. By varying the length of the alkyl chain between the $C_{60}$ core and the Hg drop, we clearly observed the reduction of the energy offset between the LUMO and the Fermi energy of the electrode when increasing the molecule-metal electronic coupling.

## 5- Experimental Section

Precursor SAMs bearing aldehyde, carbomethoxyaziridine and azide functions were prepared by classical silanization reaction. The details of these syntheses and their physicochemical characterizations are described in the supplementary information section.

### $C_{60}$-terminated SAM from $N_3$-terminated SAM.
Glassware was dried in an air oven at 120°C overnight prior to use. The reaction was performed in a reaction flask equipped with condenser and nitrogen inlet. $N_3$-terminated SAM was dipped into a degassed solution of $C_{60}$ (20 mg) in anhydrous o-dichlorobenzene (20 mL). The medium was refluxed at 180°C for 3 h under nitrogen atmosphere. After cooling, the sample was removed from the solution, sonicated for 10 min in o-dichlorobenzene then in chloroform, and finally blown with dry nitrogen.

### $C_{60}$-terminated SAM from carbomethoxyaziridine-SAM.
Carbomethoxyaziridine-terminated SAM was treated in similar conditions as $N_3$-terminated SAM. The medium was heated at 80°C for 24 h. In order to remove water traces in the solvent, the reaction was performed in the presence of freshly activated molecular sieves (4 Å).

### $C_{60}$-terminated SAM from CHO-terminated SAM (Prato reaction).
The CHO-terminated surface was immersed in a degassed solution of $C_{60}$ and N-methylglycine in anhydrous o-dichlorobenzene at reflux for 16 h. After the reaction the solution color changed from purple to dark red indicating formation of byproducts in the medium. TLC analysis revealed the presence of several adducts in the solution. Sample was removed from solution, sonicated 10 min in dichlorobenzene, chloroform, isopropanol and then dried under nitrogen.

### $C_{60}$-terminated SAM by direct grafting.
Under the same conditions as for the synthesis of $N_3$-terminated SAM, a freshly cleaned silicon substrate was immersed for 4 days in a $10^{-3}$ M solution of N-methyl-2-(3-triethoxysilylpropyl)-3,4-fulleropyrrolidine in anhydrous toluene. Sample was removed from solution, sonicated 10 min in toluene then in chloroform, then dried under nitrogen.

### Physicochemical characterizations of monolayers.
FT-IR spectra were recorded using a Perkin-Elmer Spectrum 2000 single-beam spectrophotometer equipped with a tungsten-halogen lamp and a liquid nitrogen cooled MCT detector. The functionalized silicon surfaces were characterized by reflection IR spectroscopy, method also called attenuated total reflection (ATR).[43]
Static water contact angle measurements were done using a remote computer-controlled goniometer system (DIGIDROP by GBX, France).
Spectroscopic ellipsometry data in the visible range was obtained using a UVISEL by Jobin Yvon Horiba Spectroscopic Ellipsometer equipped with a DeltaPsi 2 data analysis software. The system acquired a spectrum ranging from 2 to 4.5 eV (corresponding to 300-750 nm) with 0.05 eV (or 7.5 nm) intervals. Data were taken using an angle of incidence of 70°, and the compensator was set at 45.0°. To determine the monolayer thickness, we used the optical properties of silicon and silicon oxide from the software library, and for the monolayer we used the refractive index of 1.50. Usual values in the literature are in the range of 1.45-1.50.[18a,44] As a function of the wafer used, the oxide layer thickness was between 10 and 15 Å (accuracy ±1 Å). The $SiO_2$ thickness was assumed unchanged after monolayer assembly on the surface. Accuracy of the SAM thickness measurements is estimated to be ±2 Å. All given values for the reaction sequence were measured from a same silicon substrate.

X-ray photoemission spectroscopy (XPS) experiments were performed to control the chemical composition of the SAMs and to detect any unremoved contaminant. Samples have been analyzed by XPS, using a Physical Electronics 5600 spectrometer fitted in an UHV chamber with a base pressure of about $3.10^{-10}$ Torr. The X-ray source was monochromatic Al K$\alpha$ (h$\nu$ = 1486.6 eV) and the detection angle was 45° with respect to sample surface normal. Intensities of XPS core levels were measured as peak areas after standard background subtraction according to the Shirley procedure.

Atomic force microscope measurements were conducted in ambient air using a tapping mode AFM (Veeco, model Dimension 3100 Nanoscope).

Electrochemical investigations were carried out under nitrogen atmosphere, in anhydrous and degassed dichloromethane containing 0.1 M of tetrabutylammonium hexafluorophosphate as the supporting electrolyte. Pieces of functionalized silicon or ITO surfaces (5 mm × 10 mm) were used as working electrodes and a platinum wire was used as counter electrode. Cyclic voltammograms were recorded at 1 V.s$^{-1}$ with a potentiostat (Voltalab PGZ301) controlled by a computer. The potentials were measured *versus* a saturated calomel electrode (SCE).

etailed protocols of physicochemical characterizations of monolayers are available in the supplementary information section.

**Electrical characterizations of Si/SiO$_2$/σC$_{60}$//Hg and Si/SiO$_2$/σC$_{60}$//σHg molecular junctions.**

Measurements were performed at room temperature in a nitrogen purged glove box. Calibrated mercury drops (99.9999 %, purchased from Fluka) were generated by a controlled growth mercury electrode (CGME model from BASi). The contact area of the mercury drop with the sample surface was evaluated to $1,6 \times 10^{-3}$ cm$^2$ with a microscope camera. Si/SiO$_2$/σC$_{60}$//σHg molecular junction was prepared contacting Si/SiO$_2$/σC$_{60}$ junction with an alkanethiol-protected mercury drop (for details, see supplementary information section). Electrical transport through the SAMs was determined by measuring the current density vs the applied direct current (dc) voltage with an Agilent 4155C picoampermeter. Voltage was applied on the mercury drop, the silicon substrate being grounded.

Sequencial route 1

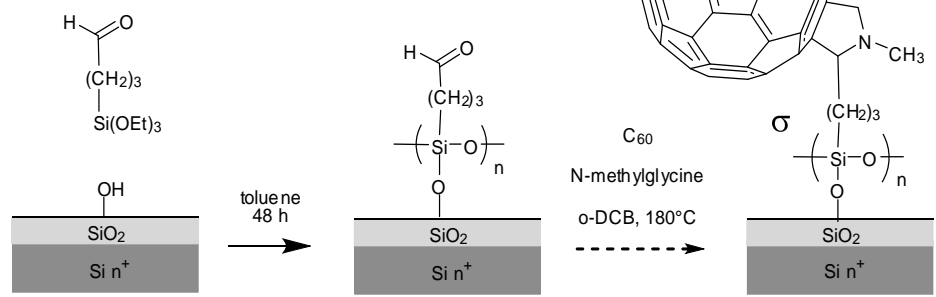

Sequencial route 2

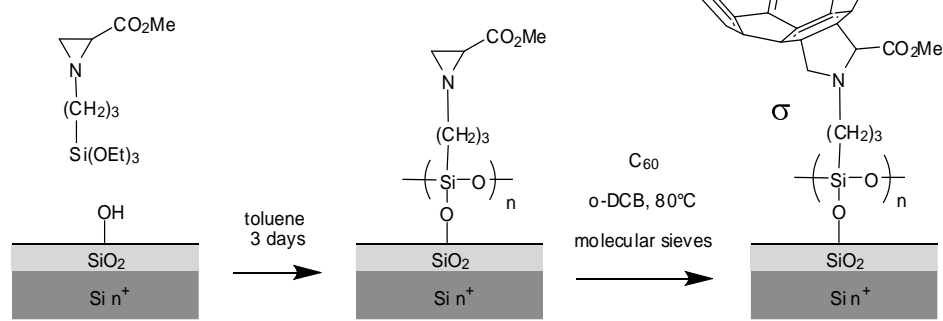

Sequencial route 3

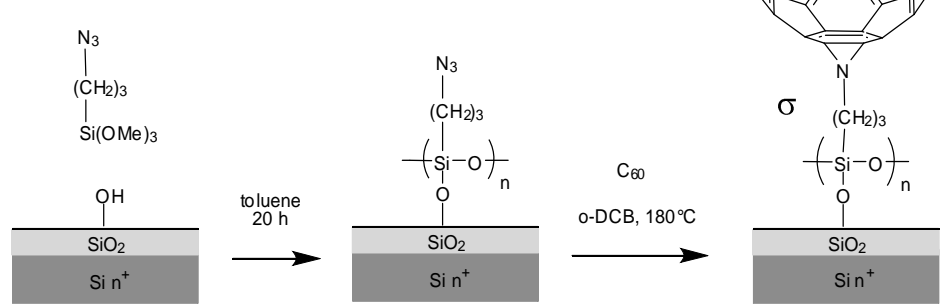

**Schema 1.** Sequential surface chemistry used to graft $C_{60}$ covalently on natively oxidized silicon

| $C_{60}$-SAM | thickness (Å) | | contact angle (°) | |
|---|---|---|---|---|
| | exp. | theo. | exp. | ref.[15b,17,22] |
| Route 1 : from CHO-SAM | 16.0 | 15.6 | 44±2 | 65-76 |
| Route 2 : from aziridine-SAM | 15.4 - 44.0 | 15.8 | 71±2 | |
| Route 3 : from $N_3$-SAM | 14.0 | 14.6 | 55 - 70±2 | |
| Direct grafting | 14.4 | 14.7 | 70±1 | |

**Table 1.** Thickness and wettability of $C_{60}$-terminated SAMs by different routes.

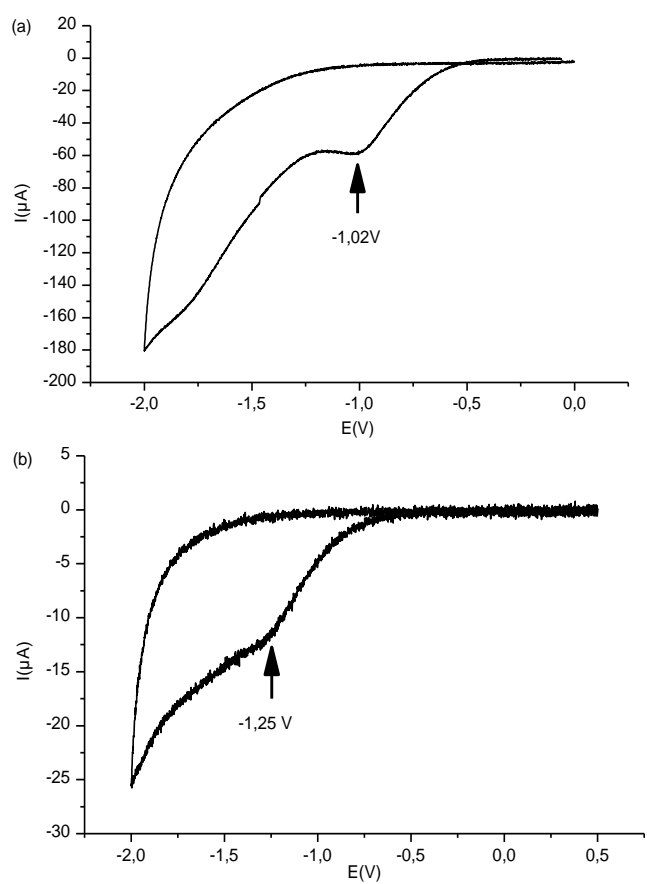

**Figure 1.** Cyclic voltammogram of $C_{60}$-terminated SAMs obtained by sequential route 2 on Si $n^+/SiO_2$. Experimental conditions : $CH_2Cl_2/NBu_4PF_6$, WE: Si $n^+/SiO_2$ (9.7Å)/$C_{60}$-SAM, CE : Pt, Ref : SCE, 1 V/s. (a) $C_{60}$-SAM thickness corresponding to 1-2 monolayers of $C_{60}$ (b) $C_{60}$-SAM thickness corresponding to 1 monolayer of $C_{60}$.

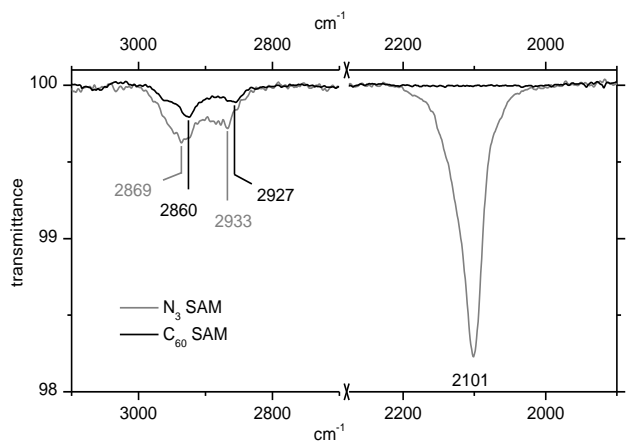

**Figure 2.** FTIR spectra of of $N_3$- and $C_{60}$-terminated SAMs (sequential route 3).

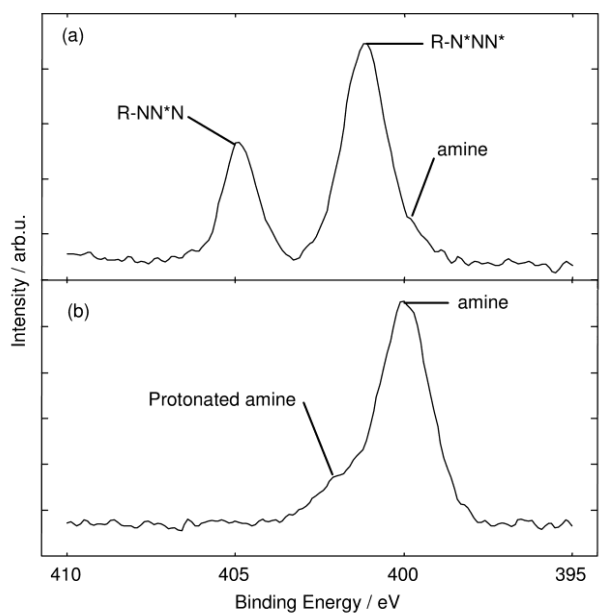

**Figure 3.** High resolution XPS spectra of the nitrogen 1s core level from (a) $N_3$- terminated and (b) $C_{60}$-terminated SAMs.

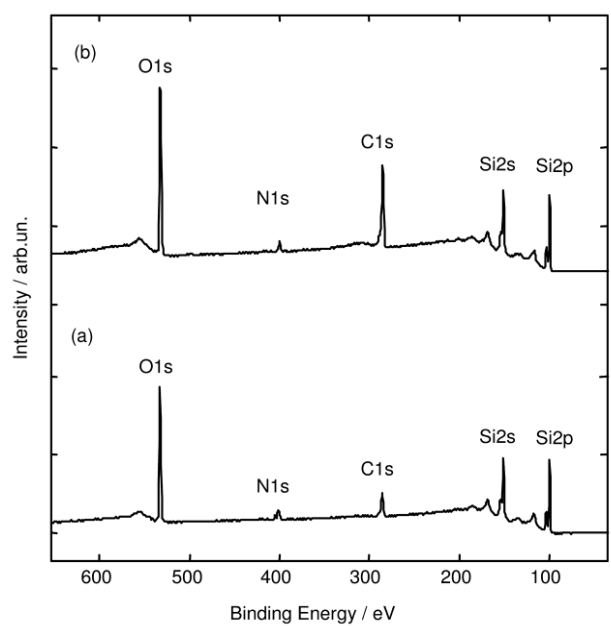

**Figure 4.** XPS survey spectra of (a) $N_3$- and (b) $C_{60}$- terminated SAMs synthesized by sequential route 3.

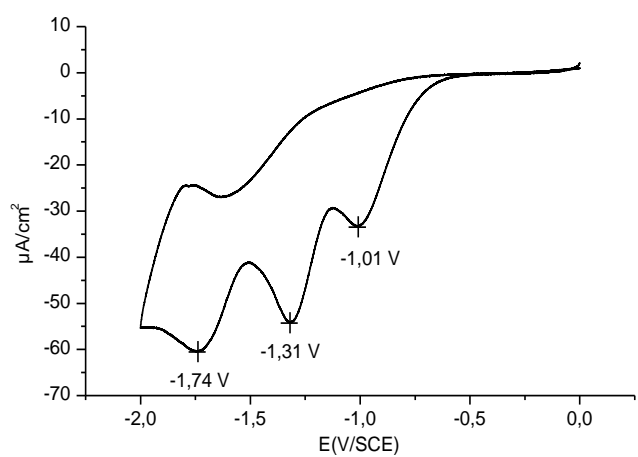

**Figure 5.** Cyclic voltammogram of $N_3$-terminated SAMs in a saturated solution of $C_{60}$ in toluene-$CH_2Cl_2$ (v/v 1:1). Experimental conditions: toluene-$CH_2Cl_2$/$NBu_4PF_6$, WE: Si $n^+$/$SiO_2$ (9.1Å)/$N_3$-SAM, CE : Pt, Ref : SCE, 10 mV/s.

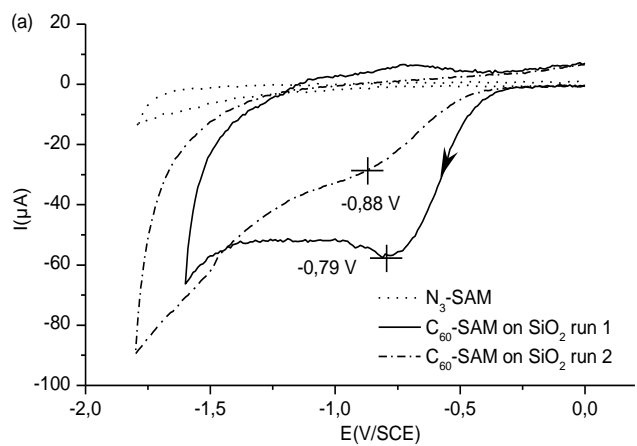

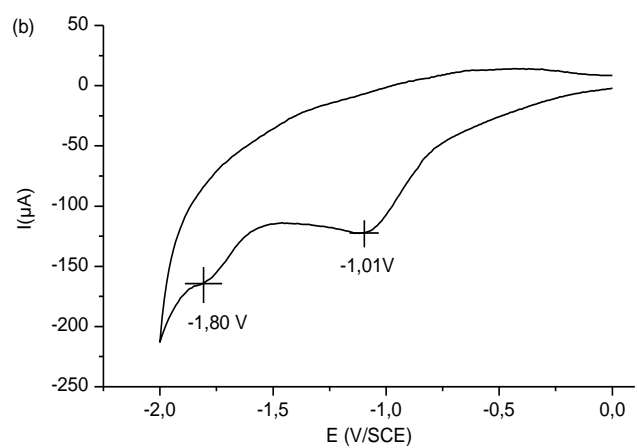

**Figure 6.** Cyclic voltammograms of $N_3$- and $C_{60}$-terminated SAMs obtained by sequential route 3 on Si n[+]/$SiO_2$ and ITO substrates. Experimental conditions : $CH_2Cl_2$/$NBu_4PF_6$, WE: Si n[+]/$SiO_2$ (9.7Å)/$C_{60}$-SAM, CE : Pt, Ref : SCE,  1 V/s. (a) $C_{60}$-SAM on Si n[+]/$SiO_2$ substrate (b) $C_{60}$-SAM on ITO substrate.

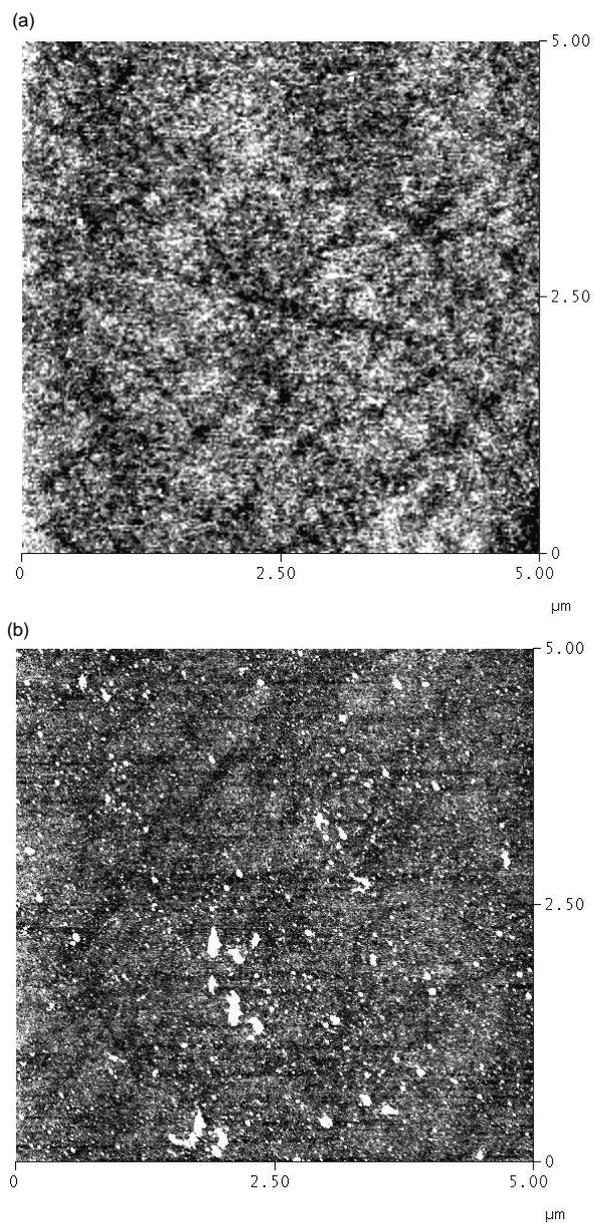

**Figure 7.** 5×5 µm AFM images of (a) $N_3$-terminated SAM and (b) $C_{60}$-terminated SAM. Diameter of $C_{60}$ clusters (white spots) was in the range 11 nm - 202 nm and their thickness was in the range 1 nm - 27 nm.

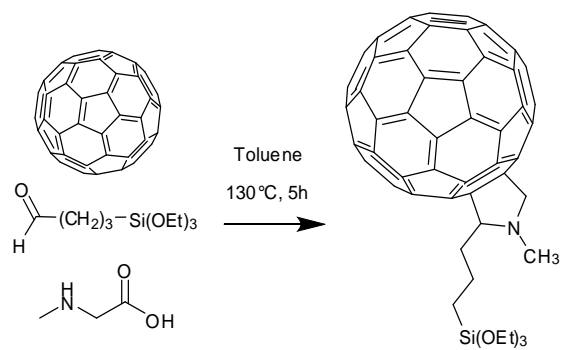

**Schema 2.** Synthesis of $C_{60}$-σ-$Si(OEt_3)$ derivative bearing anchoring group.

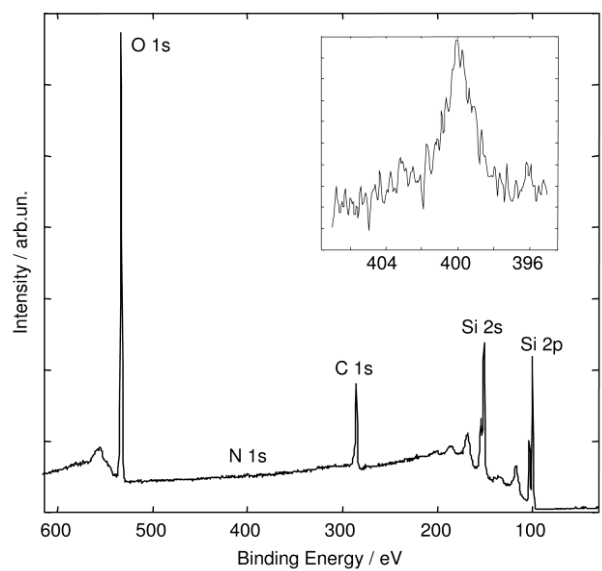

**Figure 8.** XPS survey spectrum of $C_{60}$-terminated surface obtained by direct grafting of derivative 2. The inset shows the high-resolution spectrum of N1s.

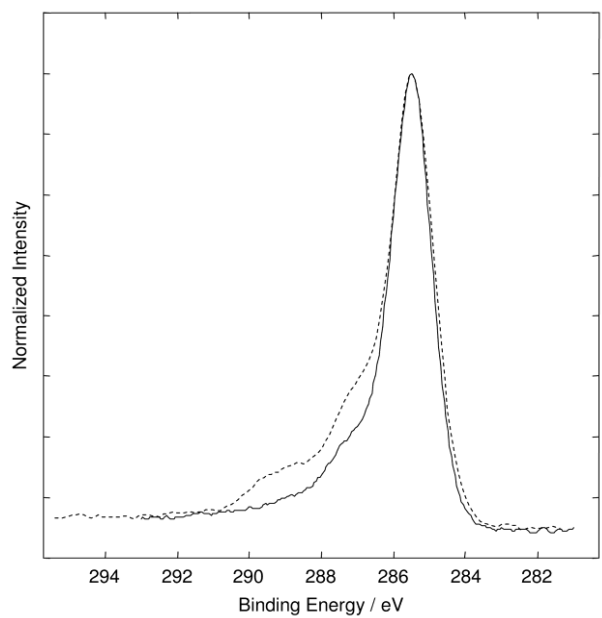

**Figure 9.** High resolution XPS spectra of the carbon 1s core level from $C_{60}$-terminated surface obtained by direct grafting (solid line) and by sequential route 3 (dashed line)

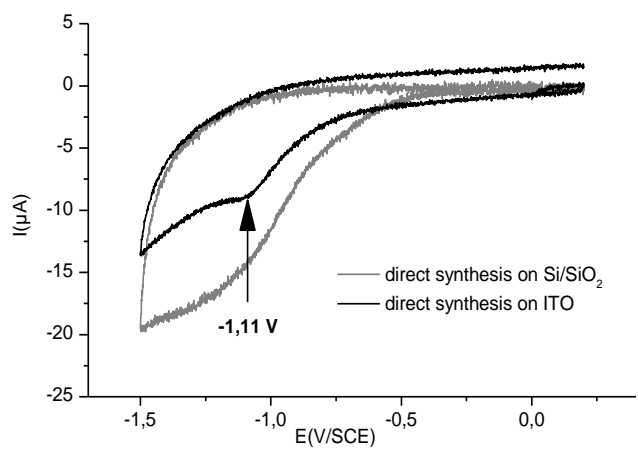

**Figure 10.** Cyclic voltammogram of $C_{60}$-terminated SAMs obtained by direct grafting on Si $n^+/SiO_2$ and ITO substrates. Experimental conditions : $CH_2Cl_2/NBu_4PF_6$, WE: Si $n^+/SiO_2/C_{60}$-SAM, CE : Pt, Ref : SCE, 1 V/s.

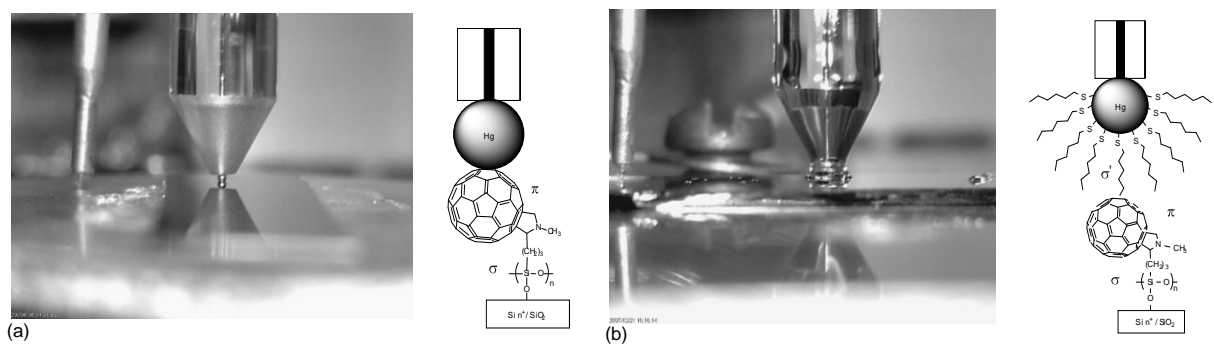

**Figure 11.** Electrical measurement of molecular junctions by mercury drop electrode. Non-thiolated (a) and thiolated (b) Hg drop in a solution of alkanethiol in hexadecane (this solvent covering the Hg drop does not evaporate and maintains a protection film during measurements).

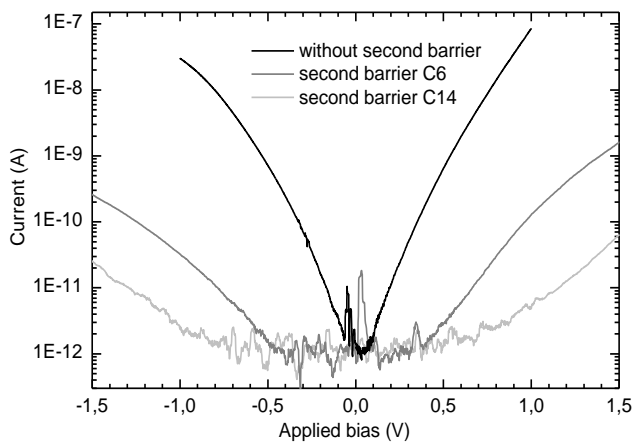

**Figure 12.** Current-voltage characteristics measured on Si/SiO$_2$/σC$_{60}$//σHg junctions for various tunnel barrier on the Hg drop : C14, C6 and without SAM. The curves were smoothed by the adjacent average method (here average with 10 neighbors). The evaluated contact area of the mercury drop with the sample surface was around $1.6 \times 10^{-3}$ cm$^2$. Voltage was applied on the mercury drop.

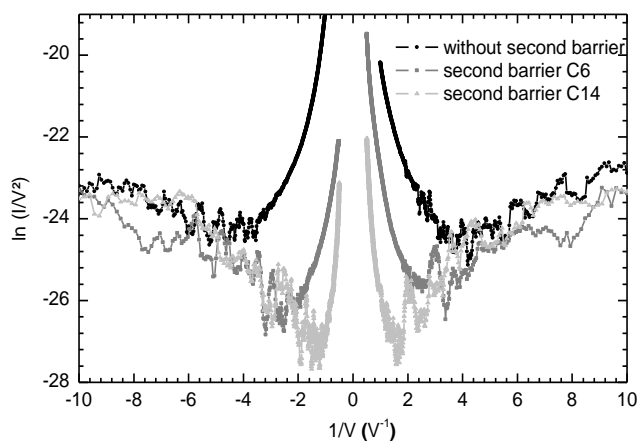

**Figure 13.** Plots of ln(I/V$^2$) versus 1/V from data in figure 12. The curves were smoothed by the adjacent average method (here average with 10 neighbors). The transition voltage V$_T$ (voltage at the minimum of the ln(I/V$^2$) determines the energy offset, $\Phi = e\,|V_T|$, between the Fermi energy of the metal electrodes and the molecular orbital involved in the charge transport through the molecular junction.

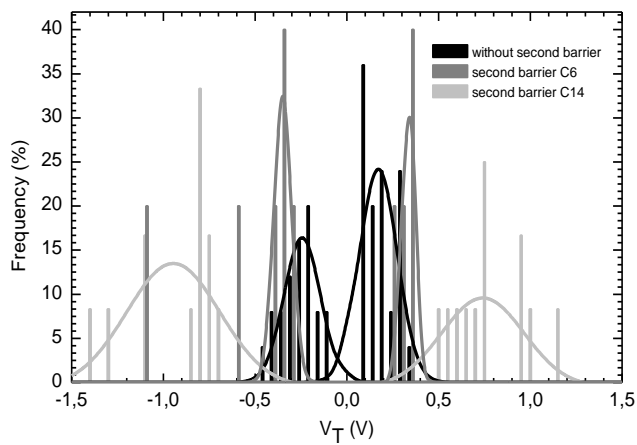

**Figure 14.** Histograms of the transition voltage $V_T$ determined from the plots of $\ln(I/V^2)$ versus $1/V$ from figure 13, for the different barrier thicknesses at negative and positive voltages. Solid lines are the fit of a normal distribution from which we deduced the mean value and standard deviation.

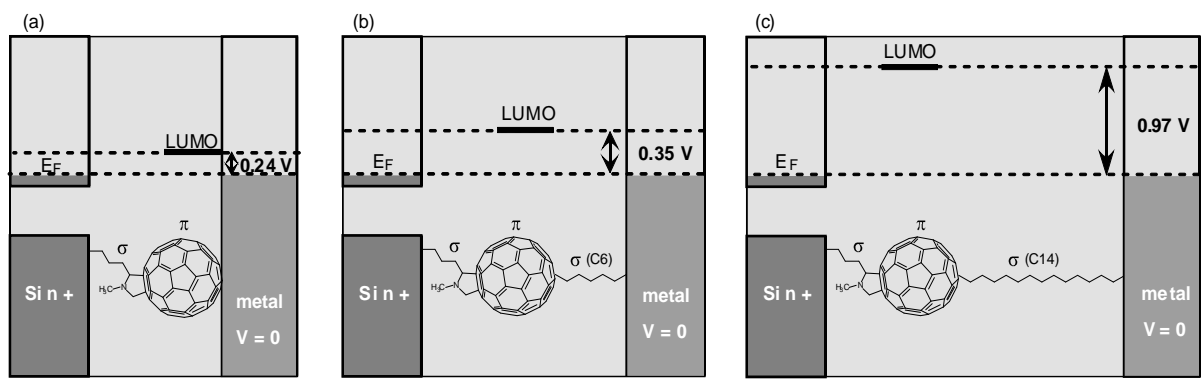

**Figure 15.** Schematic energy diagram of the Si/SiO$_2$/σC$_{60}$//σHg junctions at zero bias, determined from the TVS method, (a) without second tunnel barrier, (b) with C6 SAM for the second tunnel barrier and (c) with C14 SAM for the second tunnel barrier. For clarity, we do not show the native SiO$_2$ (acting as an other tunnel barrier in series with short C$_3$ alkyl chain attaching the C$_{60}$ on the substrate, and we neglect the work function difference between the Hg electrode and the highly doped silicon substrate n$^+$-type Si (4.5eV for clean Hg and 4.1eV for n$^+$-type Si)

# Supporting information

# Synthesis and electrical properties of fullerene-based molecular junctions on silicon substrate

D. Guérin, S. Lenfant, S. Godey, D. Vuillaume

## 1- Synthesis of organosilanes

**Chemicals.**
Buckminsterfullerene $C_{60}$ (sublimed, 99.9%), N-methylglycine (99 %), sulfuric acid (96 %), hydrogen peroxide (30 % in water), sodium azide (99.5 %), adogen 464, tetrabutylammonium hexafluorophosphate (99 %), anhydrous o-dichlorobenzene (99 %) were all purchased from Sigma Aldrich. N-[3-(triethoxysilyl)propyl]-2-carbomethoxyaziridine (97 %), triethoxysilylbutyraldehyde (90 %), 3-chloropropyltrimethoxysilane (95 %) were purchased from Gelest. Chloroform (99.9 %) and dichloromethane (99.9 %) were obtained from Scharlau. Anhydrous acetonitrile (99.8 %), toluene (99.9 %), ethyl acetate (99.5 %), isopropanol (99.7 %) were purchased from Carlo Erba. These chemicals were used as received without further purification, except dichloromethane and toluene that were distilled over calcium hydride then stored over freshly activated molecular sieves. For water-sensitive reactions, glassware was dried in an oven at 120 °C overnight prior to use.

**$^1$H and $^{13}$C NMR spectra** were recorded with AC 300 Bruker instrument. $^1$H NMR spectra were recorded at 300 MHz and $^{13}$C NMR spectra at 75 MHz. Chemical shifts of NMR spectra are reported in ppm and referenced to tetramethylsilane.

**FTIR spectroscopy** of synthesized 3-azidopropyltrimethoxysilane was realized using a Perkin-Elmer Spectrum 2000 single-beam spectrophotometer by transmission IR spectroscopy in a sodium chloride crystal cell.

**3-azidopropyltrimethoxysilane**
3-azidopropyltrimethoxysilane was synthesized following the procedure developed by Karl et al.[1] by reaction of sodium azide with 3-chloropropyltrimethoxysilane in the presence of Adogen 464 catalyst. Let us precise that $N_3$-terminated SAMs is often also prepared by surface chemistry from halide-terminated monolayers[2,3] but this method would have required one more chemical step in the reaction sequence and thus one more risk of pollution and decrease of yield.
All glassware was dried in an air oven at 120 °C overnight prior to use and the reaction was performed under nitrogen atmosphere. A solution of 3-chloropropyltrimethoxysilane (2.5 mL, 13.71 mmol), sodium azide (1 g, 15.4 mmol) in dry acetonitrile was refluxed for 16 h under stirring in the presence of the catalyst Adogen 464 (200 mg, 0.4 mmol). After filtration of the precipitated salt, then evaporation of the solvent, the resulting oil was purified by distillation under reduced pressure (0.5 Torr). The first fraction (below 40°C) containing the unreacted halogenated precursor was eliminated then 3-azidopropyltrimethoxysilane was obtained at 80°C as a clear liquid in a yield of 60 % (1.7 g). *Caution: due to the excess of instable and highly toxic sodium azide, the oil should not be distilled to dryness.* $^1$H NMR (300 MHz; CDCl$_3$; Me$_4$Si; 20°C) : $\delta_H$(ppm) 0.67 (2H, m, -CH$_2$Si), 1.68 (2H, m, -CH$_2$CH$_2$CH$_2$-), 3.24 (2H, t, $^3J_{CH}$ = 6.95 Hz, -CH$_2$N$_3$), 3.55 (9H, s, CH$_3$O). $^{13}$C NMR (75.5 MHz; CDCl$_3$; Me$_4$Si; 20°C) : $\delta_C$(ppm) 6.27, 22.41, 50.50, 53.67. FTIR (in CCl$_4$) : $\lambda_{max}$ 2098.8 cm$^{-1}$ (-N$_3$).

**N-methyl-2-(3-triethoxysilylpropyl)-3,4-fulleropyrrolidine**

A solution of C$_{60}$ (100 mg, 139 μmol), N-methylglycine (50 mg, 561 μmol), and triethoxysilylbutyraldehyde (150 μL, 553 μmol) in toluene (50 mL) was heated to reflux for 5 h. The solvent was evaporated then the crude product was purified by flash chromatography (SiO$_2$, 40-63 μm) with toluene first to remove the unreacted C$_{60}$, then with a mixture of toluene/ethyl acetate v/v 9:1. The fulleropyrrolidine was isolated as a brown solid (32 mg, 22 %) by filtration after dissolution in a minimum of toluene then precipitation by the addition of acetonitrile. During the flash chromatography, 50 % of unreacted fullerene was also recovered. R$_f$ = 0.47 in toluene/ethyl acetate, v/v 9:1. $^1$H NMR (300 MHz; C$_6$D$_6$; Me$_4$Si; 20 °C): $\delta_H$(ppm) 0.85 (2 H, t, CH$_2$Si), 1.18 (9 H, t, J = 6.9 Hz, CH$_3$CH$_2$O), 1.34-1.58 (4H, m, CH$_2$CH$_2$CH$_2$Si), 2.65 (3H, s, CH$_3$N), 3.80 (6H, q, J = 6.9 Hz, CH$_3$CH$_2$O), 3.74 (1H, t, J = 5.55 Hz, CHCH$_2$), 3.87-3.92 (1H, m, CH$_2$ pyrrolidine), 4.34 (1H, d, J = 9.78 Hz, CH$_2$ pyrrolidine). $^{13}$C NMR (75.5 MHz; C$_6$D$_6$; Me$_4$Si; 20 °C): $\delta_C$(ppm) 11.77, 15.71, 21.48, 34.62, 39.48, 58.47, 70.22, 70.76, 76.85, 78.11, 135.91, 136.31, 136.71, 137.66, 140.01, 140.14, 140.51, 140.58, 142.02, 142.06, 142.19, 142.39, 142.42, 142.46, 142.49, 142.50, 142.56, 142.58, 142.98, 143.01, 143.05, 143.44, 143.56, 144.72, 144.78, 144.94, 145.12, 145.51, 145.54, 145.57, 145.60, 145.65, 145.75, 145.77, 145.93, 145.94, 146.16, 146.26, 146.28, 146.36, 146.40, 146.52, 146.53, 146.59, 146.63, 146.75, 147.06, 147.31, 147.48, 147.52, 154.34, 154.95, 155.05, 157.19.

**2- Synthesis of aldehyde, carbomethoxyaziridine and azide terminated SAMs**

CHO-terminated SAM, carbomethoxyaziridine-terminated SAM and N$_3$-terminated SAM were successfully prepared by classical silanization reactions immersing a freshly cleaned natively oxidized silicon substrate in a 10$^{-3}$ M solution of the corresponding organosilane in toluene for several days.

**Silicon substrate cleaning method.** Substrates used for electrical characterizations were n-type degenerated silicon wafers purchased from Siltronix (resistivity of ~10$^{-3}$ Ω.cm) to avoid any voltage drop during electrical measurements. For FTIR spectroscopy, ATR silicon crystals (undoped) were used (45°, 10 mm × 5 mm × 1.5 mm). Surface of silicon substrates were covered with native oxide (10-15 Å thick) providing a dense array (~5 × 10$^{14}$ cm$^{-2}$) of reactive hydroxyl groups (-OH), which are anchoring sites for the alkyltrialkoxysilane molecules. All silicon substrates including ATR silicon crystals were first sonicated in chloroform for 5 min then dried under nitrogen. The polished side of the wafers was cleaned by ozonolysis for 30 min in a UV/ozone cleaner (wavelength at 185 and 254 nm). Then, the samples were dipped into a freshly prepared piranha solution (H$_2$SO$_4$-H$_2$O$_2$ 2:1 v/v) at 100 °C for 15 min, were rinsed thoroughly with DI water then were dried under nitrogen stream. Finally, cleaning was ended by additional ozonolysis for 30 min. *Caution: the piranha solution violently reacts with organic chemicals; consequently, it should be handled with extreme care.*

**N$_3$-terminated SAM**
The reaction was carried out at room temperature in a glovebox under nitrogen atmosphere. A freshly cleaned silicon substrate was immersed for 48 h in a $10^{-3}$ M solution of 3-azidopropyltrimethoxysilane in anhydrous toluene (i.e. ~20 µL of organosilane in 20 mL of solvent). The substrate was removed, sonicated 5 min in toluene then in chloroform, and blown with dry nitrogen.

**Carbomethoxyaziridine-terminated SAM**
The SAM was prepared by treating the freshly oxidized silicon substrate with N-[3-(triethoxysilyl)propyl]-2-carbomethoxyaziridine in the same conditions as N$_3$-SAM. The immersion duration in toluene was 3 days.

**CHO-terminated SAM**
Freshly cleaned silicon substrate was immersed for 3 days in a $10^{-3}$ M solution of triethoxysilylbutyraldehyde in anhydrous toluene in the same conditions as N$_3$-SAM.

As can be seen in table 1, precursor SAMs thicknesses measured by ellipsometry were conform to theoretical values (MOPAC, PM3 optimization) calculated considering the molecules in their all-trans conformation and almost standing perpendicular to the surface. Measured water contact angles (see table 1) were slightly lower than those of analogous monolayers published in literature. In the present work, the molecule lengths are much smaller than those previously described, and it is known[4] that SAM of short chains are more disordered than long ones, thus explaining the lower contact angles.

| Precursor SAMs | thickness (Å) | | contact angle (°) | |
|---|---|---|---|---|
| | exp. | theo. | exp. | ref. |
| CHO-SAM | 7.0 | 7.8 | 67±1 | 80[5] |
| Aziridine-SAM | 12.0 | 12.1 | 55±1 | NR |
| N$_3$-SAM | 8.0 | 8.8 | 68±1 | 77[6] |

**Table 1.** Thickness and wettability of precursor SAMs selected for sequential methods. NR : no found reference in literature.

**3- Instrumentation and methods for the physicochemical characterizations of self-assembled monolayers**

**FTIR spectroscopy.** FT-IR spectra were recorded using a Perkin-Elmer Spectrum 2000 single-beam spectrophotometer equipped with a tungsten-halogen lamp and a liquid nitrogen cooled MCT detector. The functionalized silicon surfaces were characterized by reflection IR spectroscopy, method also called attenuated total reflection (ATR).[7] For the IR studies, all substrates were ATR silicon crystals (45°, 10 mm×5 mm×1.5 mm). All measurements were made after purging the sample chamber for 30 min with dry N$_2$. Spectra were recorded at 4 cm$^{-1}$ resolution, and 200 scans were averaged. Background spectra of the freshly cleaned crystal (with SiO$_2$ layer) were recorded before any chemical treatment. After the chemical reactions, the sample spectra were recorded in the same conditions. The monolayer transmittance spectrum was obtained by dividing the sample spectrum by the background spectrum.

**Contact angle measurements** were done using a remote computer-controlled goniometer system (DIGIDROP by GBX, France). These measurements were carried out in a clean room (class 1,000) where humidity, temperature, and pressure were controlled. Deionized water (18 MΩ/cm) was used for these characterizations. Static drops (in the range 1-10 µL) of liquid were applied to the modified surfaces with a micropipette and the projected image was acquired and stored by the remote computer. Several measurements were taken on each region of a given substrate. Contact angles were then extracted by contrast contour image analysis software. These angles were determined 5 s after application of the drop. Measurements made across the functionalized surfaces were within 2°.

**Spectroscopic ellipsometry** data in the visible range was obtained using a UVISEL by Jobin Yvon Horiba Spectroscopic Ellipsometer equipped with a DeltaPsi 2 data analysis software. The system acquired a spectrum ranging from 2 to 4.5 eV (corresponding to 300-750 nm) with 0.05 eV (or 7.5 nm) intervals. Data were taken using an angle of incidence of 70°, and the compensator was set at 45.0°. Data were fitted by regression analysis to a film-onsubstrate model as described by their thickness and their complex refractive indices. First, before monolayer deposition the silicon oxide thickness was measured. In the software, the measured data were compared with the simulated data to determine this thickness. The simulated data were obtained with a 2-layer model ($Si/SiO_2$). In this model we use for the materials, the optical properties (complex refractive index for each wavelength) from the software library. Second, after the monolayer deposition we use a 3-layer model ($Si/SiO_2$/organic film). To determine the monolayer thickness, we use the oxide thickness previously measured, for silicon and oxide we use the optical properties from the software library, and for the monolayer we use the refractive index of 1.50. Usual values in the literature are in the range of 1.45-1.50.[8,9] One can notice that a change from 1.50 to 1.55 would result in less than 1 Å error for a thickness less than 30 Å. As a function of the wafer used, the oxide layer thickness was between 10 and 15 Å (accuracy ±1 Å). The $SiO_2$ thickness is assumed unchanged after monolayer assembly on the surface. Accuracy of the SAM thickness measurements is estimated to be ±2 Å. All given values for the reaction sequence were measured from a same silicon substrate.

**X-ray photoemission spectroscopy (XPS)** experiments were performed to control the chemical composition of the SAMs and to detect any unremoved contaminant. Samples have been analyzed by XPS, using a Physical Electronics 5600 spectrometer fitted in an UHV chamber with a base pressure of about $3.10^{-10}$ Torr. The X-ray source was monochromatic Al Kα (hν = 1486.6 eV) and the detection angle was 45° with respect to sample surface normal. Intensities of XPS core levels were measured as peak areas after standard background subtraction according to the Shirley procedure.

**Atomic force microscope measurements** were conducted in ambient air using a tapping mode AFM (Veeco, model Dimension 3100 Nanoscope) equipped with a silicon cantilever (tip radius <20 nm).

**Electrochemical investigations** were carried out under nitrogen atmosphere, in anhydrous and degassed dichloromethane containing 0.1 M of tetrabutylammonium hexafluorophosphate as the supporting electrolyte. Pieces of functionalized silicon surfaces (5 mm × 10 mm) were used as working electrodes and a platinum wire was used as counter electrode. Cyclic voltammograms were recorded at 1 $V.s^{-1}$ with a potentiostat (Voltalab PGZ301) controlled by a computer. The potentials were measured *versus* a saturated calomel electrode (SCE). For sequential route 3 and direct grafting, indium tin oxide substrates (ITO) were employed as working electrodes instead of Si/SiO$_2$ in order to amplify electrochemical signal. ITO was first cleaned by sonication in chloroform for 5 min then treated by UV/ozone cleaner for 15 min before being functionalized exactly in the same conditions as Si/SiO$_2$.

## 4- Experimental procedure for electrical characterizations of Si/SiO$_2$/σC$_{60}$//Hg and Si/SiO$_2$/σC$_{60}$//σHg molecular junctions

Measurements were performed at room temperature in a nitrogen purged glove box. The Si n$^+$/SiO$_2$/σC$_{60}$ junctions were fixed onto copper substrates with silver paste. The wafer backside was stripped with a diamond tip to remove backside oxide and ensure a good electrical contact with the silver paste. These copper substrates were mounted onto a precision vertical translation stage and the top contact was realized by carefully approaching the sample surface towards a hanging mercury drop electrode. Mercury (99.9999 %) was purchased from Fluka. Calibrated mercury drops were generated by a controlled growth mercury electrode (CGME model from BASi). The contact area of the mercury drop with the sample surface was evaluated to $1,6 \times 10^{-3}$ cm$^2$ with a microscope camera. Si/SiO$_2$/σC$_{60}$//σHg molecular junction was prepared contacting Si/SiO$_2$/σC$_{60}$ junction with an alkanethiol-protected mercury drop. This treatment was obtained by immersing a hanging mercury drop in a 10$^{-3}$ M solution of hexadecanethiol (C6) or tetradecanethiol (C14) in hexadecane for 18 h. The thiolated drop was removed from the solution then directly applied on the Si/SiO$_2$/C$_{60}$ junction (see fig. 11). During measurements, the mercury drop electrode is protected by a film of the alkylthiol solution in hexadecane (this solvent does not evaporate). Electrical transport through the SAMs was determined by measuring the current density vs the applied direct current (dc) voltage with an Agilent 4155C picoampermeter. Voltage was applied on the mercury drop, the silicon substrate being grounded.